\documentclass{jetpl}
\twocolumn

\usepackage{amsmath,amssymb,epsfig,color,pstricks,bm,graphics,alltt}

\begin{document}

\lat

\title{Comparative Study of Electronic structure of New Superconductors
(Sr,Ca)Pd$_2$As$_2$ and related compound BaPd$_2$As$_2$.}

\rtitle{Electronic structure: (Sr,Ca)Pd$_2$As$_2$ vs. BaPd$_2$As$_2$}

\sodtitle{Comparative Study of Electronic structure of New Superconductors
(Sr,Ca)Pd$_2$As$_2$ and related compound BaPd$_2$As$_2$.}

\author{$^{a}$I.\ A.\ Nekrasov\thanks{E-mail: nekrasov@iep.uran.ru},
$^{a,b}$M.\ V.\ Sadovskii\thanks{E-mail: sadovski@iep.uran.ru}}

\rauthor{I.\ A.\ Nekrasov, M.\ V.\ Sadovskii}

\sodauthor{Nekrasov, Sadovskii}

\sodauthor{Nekrasov, Sadovskii}

\address{$^a$Institute for Electrophysics, Russian Academy of Sciences, 
Ural Branch, Amundsen str. 106,  Ekaterinburg, 620016, Russia\\
$^b$Institute for Metal Physics, Russian Academy of Sciences, Ural Branch,
S.Kovalevskoi str. 18, Ekaterinburg, 620990, Russia}


\abstract{
This paper presents the comparative study of LDA calculated electronic structure
of new isostructural to iron based systems superconductors (Sr,Ca)Pd$_2$As$_2$
with T$_c$ about 1K and similar but structurally different system BaPd$_2$As$_2$.
Despite chemical formula looks similar to iron superconductors and
even main structural motif is the same - layers of Fe square lattices,
electronic structure of (Sr,Ca)Pd$_2$As$_2$ and BaPd$_2$As$_2$
differs from Fe(As,Se)-HTSC completely.
All these systems have essentially three dimensional Fermi surfaces in contrast to Fe(As,Se) materials.
The Fermi level is crossed by low intensive tails of Pd-4d and As-4p states.
However (Sr,Ca)Pd$_2$As$_2$ and BaPd$_2$As$_2$ materials have
rather well developed peaks of Pd-4d($x^2-y^2$) band. Thus by doping
of about 2 holes per unit cell one can increase density of states
at the Fermi level by a factor about 2.5. Since experimentally
these compounds were found to be simple BCS superconductors
the hole doping may considerably increase T$_c$.
LDA calculated total densities of states at the Fermi level 
for stoichiometric systems perfectly agree with experimental
estimates signifying rather small role of electronic correlations.}

\maketitle

Novel noncuprate HTSC FeAs based systems after their discovery in 2008 
\cite{kamihara_08} stimulated an avalanche of experimental and theoretical 
investigations~\cite{UFN_90,Hoso_09,MKrev}. Moreover long series of
related perspective compounds was synthesized.

Electronic stucture of many  FeAs systems was investigated by 
us \cite{Nekr,Nekr2,Nekr3,Nekr4,Kucinskii10} and other groups \cite{Shein_Fe,Singh_Fe}.
Also we made some effort to investigate FeSe based superconductors
(chalcogenides) \cite{Singh_Se,kfese_Nekr,kfese_Shein,Craco_Se} and to compare them with pnictides \cite{PvsC}.
Electronic structures of some related systems were studied during recent years:
APt$_3$P \cite{apt3p}, BaFe$_2$Se$_3$ \cite{Ba123}, SrPt$_2$As$_2$ \cite{srpt2as2}.

This work was motivated by recent detailed work of Anand $et~al.$~\cite{Anand}
on superconducting and normal state properties of (Sr,Ca)Pd$_2$As$_2$ and BaPd$_2$As$_2$
single crystals. SrPd$_2$As$_2$ and CaPd$_2$As$_2$ systems were found to be superconductors
with T$_c$ 0.92K and 1.27K correspondingly, while the BaPd$_2$As$_2$ showed only
traces of superconductivity. These Pd systems
can be considered as a physically interesting end-point compounds of A(Fe$_{1-x}$M$_x$)$_2$As$_2$
with M=Cr,Mn,Co,Ni,Cu,Ru,Rh and A=Ca,Sr,Ba series
with respect to maximal number of charge carriers in such systems.
Changes of magnetic and superconducting properties
of A(Fe$_{1-x}$M$_x$)$_2$As$_2$ series are quite non trivial.
Detailed overview of electronic properties of these compounds is presented in the
paper by Anand $et~al.$~\cite{Anand}.
However theoretical band structures of SrPd$_2$As$_2$, CaPd$_2$As$_2$ and BaPd$_2$As$_2$
materials were not investigated yet to our knowledge. 

Here we present the detailed comparison of band structures of
SrPd$_2$As$_2$, CaPd$_2$As$_2$ and BaPd$_2$As$_2$ materials with respect to each other and
also to isovalent  (Sr,Ba)Ni$_2$As$_2$ compounds \cite{Subedi_Ni,Zhou_Ni,Shein_Ni}
 and iron pnictides and chalcogenides in general.

For our band structure calculations within the local density approximation (LDA)
we used refined crystal structure data of Ref.~\cite{Anand}. SrPd$_2$As$_2$ and CaPd$_2$As$_2$
have tetragonal body centred crystal structure with the space group $I/4mmm$
the same as typical pnictide representative BaFe$_2$As$_2$ \cite{rott} where two mirrored 
FeAs$_4$ tetrahedra layers are contained in the elementary cell \cite{Nekr2}
(see also Fig.~1 on the left side).
Although in most cases 122-pnictide systems belong to the space group $P4/nmm$ 
\cite{Kucinskii10}. The lattice parameters are $a$=4.2824\AA~and 4.3759\AA~
and $c$=10.088\AA~and 10.1671\AA~for Ca and Sr systems respectively.
As compared to Ba122 pnictide the Ca and Sr systems
have larger $a$ and smaller $c$ parameters.
Wyckoff positions of ions are identical to 122 systems:
Ca,Sr -- 2a(0,0,0); Pd -- 4d(0,0.5,0.25); As -- 4e 
(0,0,0.3763) for Ca and (0,0,0.3768) for Sr systems.
\begin{figure}[ht]
\begin{center}
\includegraphics[clip=true,width=0.22\textwidth]{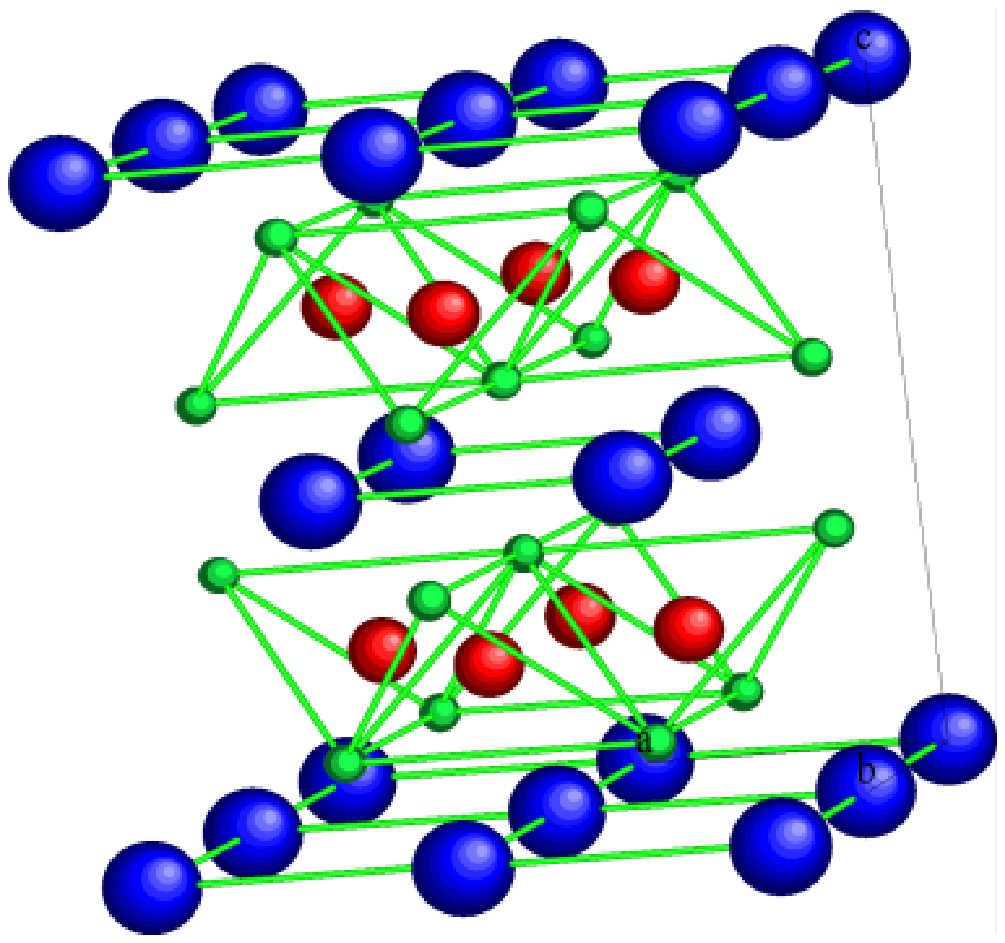}
\includegraphics[clip=true,width=0.2\textwidth]{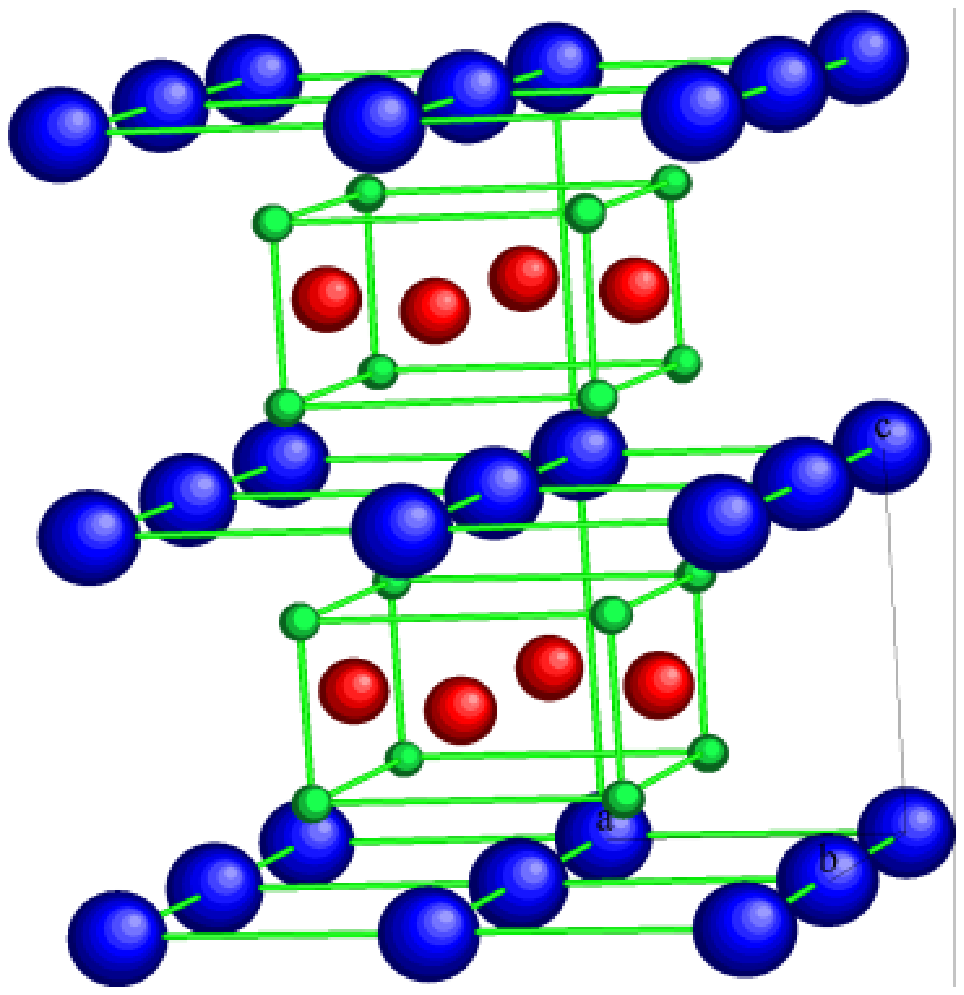}
\label{fig1}
\end{center}
\caption{Fig. 1. Crystal structure of (Sr,Ca)Pd$_2$As$_2$ (left) and BaPd$_2$As$_2$ (right).
Blue balls are Sr,Ba ions, green -- As and red -- Pd.} 
\end{figure}

Being chemically similar to Ba122 pnictide the BaPd$_2$As$_2$
system has very different crystal structure. According to Ref.~\cite{Anand}
the space group of BaPd$_2$As$_2$ is $P/4mmm$. Ions  positions are:
Ba -- 1a(0,0,0); Pd -- 2e(0,0.5,0.5); As -- 2h(0.5,0.5,0.2505).
This crystal structure is drawn in Fig.~1 on the right side.
The crystal structure also consists from layers.
But in contrast to Ba122 and other Fe-pnictides and chalcogenides
Fe ions here (which form square lattice) are surrounded by
As rectangles (see the right panel of Fig.~1) and not tetrahedrons
(e.g. see Fig.~1, left panel).

Electronic structure calculations were performed within the linearized muffin-tin orbitals 
method (LMTO)~\cite{LMTO} with default settings.
LDA calculated  band dispersions plotted along high-symmetry Brillouin zone directions
are shown in Fig.~2 for SrPd$_2$As$_2$ (top panel) and BaPd$_2$As$_2$ (bottom panel) on the right sides.
Since band structure of CaPd$_2$As$_2$ obtained by us is only slightly different
from that of SrPd$_2$As$_2$ below we present only theoretical data for SrPd$_2$As$_2$ system.

For the Sr system our band dispersions are quite similar to
those obtained in Ref.~\cite{Shein_Ni} for isovalent SrNi$_2$As$_2$ compound
except a little bit shifted up in energy part of As-4p states within X-P direction.
This leads to a slightly simpler than in Ni case Fermi surface, which is plotted in Fig.~4 (upper row)
and described below. Around $\Gamma$ we observe small hole pocket while
around X-point we see large electronic pocket. Close to N-point
there is another electronic pocket of the Fermi surface.
Fermi level is crossed by bands containing many contributions of different states
without any dominant orbital.

It is interesting that band dispersions and Fermi surfaces of isovalent BaNi$_2$As$_2$ compound 
reported in Refs.~\cite{Subedi_Ni,Zhou_Ni} are quite different from those of SrNi$_2$As$_2$ and SrPd$_2$As$_2$.
Which agrees with our calculations.
This is rather puzzling since BaNi$_2$As$_2$ crystal structure is very much similar to
SrNi$_2$As$_2$ and SrPd$_2$As$_2$ compounds.
At the same time densities of states of BaNi$_2$As$_2$ are rather close to those
plotted on the right side of upper panel of Fig.~2 for SrPd$_2$As$_2$.

From SrPd$_2$As$_2$ density of states 
one can see that most of the spectral weight is formed by Pd-4d and As-4p states.
Pd-4d states are located between -4 and -0.5 eV (see also the upper panel of Fig.~3), As-4p states
belong to the interval (-6;-4) eV. Significant hybridization between
Pd-4d and As-4p states is also evident. Comparing this Sr system to e.g. Ba122 Fe-pnictide
we see that Pd-4d states are obviously more extended in energy than Fe-3d.
Overall shape of the SrPd$_2$As$_2$ total DOS as expected is similar to that of SrNi$_2$As$_2$ and to Ba122 Fe-pnictide.
However the value of total DOS at the Fermi level $N(E_F)$=1.93 states/eV/f.u. is more than twice
lower than for Ba122 Fe-pnictide because additional electron doping moves the Fermi level up into lower 
DOS region. Indirect estimates of $N(E_F)$ from experiments
gives 1.89 states/eV/f.u. \cite{Anand} which agrees well with our calculated value,
signifying that correlation effects are more or less not important in the SrPd$_2$As$_2$.
This can be due to more extended in energy Pd-4d states in contrast to Fe-3d states e.g. in Ba122 Fe-pnictide.

To understand the structure of Pd-4d DOS of the SrPd$_2$As$_2$ we
present (on the upper panel of Fig.~3) the orbitally resolved DOSes.
Here we see that about 0.8eV below the Fermi level there is
peak formed by Pd-4d($x^2-y^2$). Thus by doping of about
2 holes per unit cell one can increase $N(E_F)$ more than twice.
Consequently one can expect increase of the $T_c$ since
this system was experimentally found to be a simple BCS like
superconductor \cite{Anand}.

Lower panel of Fig.~2 demonstrates LDA electronic band dispersions
and densities of states of BaPd$_2$As$_2$. Since its crystal structure
is totally different from all other Fe pnictides and chalcogenides
no wonder that its electronic structure is different too.
Around $\Gamma$ and R points we found large and small hole-like
pockets and around M point there are two electron-like pockets
(see also lower line of the Fig.~4).
Similar to the Sr material in case of Ba compound Fermi level is also crossed by bands
containing many contributions of different states
without any dominant orbital contribution and finally giving quite low DOS.

Despite different band dispersions, the shape of total and partial DOSes
for Ba compound is somewhat similar to Sr ones. Pd-4d states are located between
-3 eV and -0.5 eV, while As-4p states are situated right below Pd-4d states
and continue down to -6 eV. Considerable admixture (hybridisation)
between Pd-4d and As-4p states is observed.
The LDA calculated value of total density of states at the Fermi level $N(E_F)$
for BaPd$_2$As$_2$ is 2.29 states/eV/f.u. (taking into account the fact that there is
only one f.u. in the unit cell for Ba material). This value agrees well
with experimental estimates of Ref.~\cite{Anand} giving 2.03 states/eV/f.u..
However one can guess that electron-phonon coupling constant obtained experimentally
in Ref. \cite{Anand} is slightly overestimated or some other rather weak interactions
should be taken into account.

Similar to the Sr system in BaPd$_2$As$_2$ about 0.7eV below the Fermi level
there is peak in the Pd-4d DOS due to the band of $x^2-y^2$ symmetry
which is clearly seen at this energy within dispersions.
Thus we also can expect here that hole doping can lead to appearance of
superconductivity (its traces were actually observed experimentally in Ref.~\cite{Anand}).

In Fig.~4 we present Fermi surfaces of SrPd$_2$As$_2$ (upper row) and
BaPd$_2$As$_2$ (lower row) obtained within LDA calculations.
For both compounds panels (a) give general view of the Fermi surfaces
in corresponding first Brillouin zone. Fermi surface of the Sr system is
essentially three-dimensional in contrast to e.g. 122 Fe-pnictides (see Ref.~\cite{Nekr2}).
The structure of the Fermi surface is quite complicated and
consists of three sheets. On the panel (b) (upper row of Fig.~4)
there is hole-like sheet in according to performed above band structure
analysis but this sheet does not cross $k_z=0$ plane. On panels (c) and (d)
hole and electronic sheets are presented. The latter one cross
the  $k_z=0$ plane forming the Fermi surface presented on the panel (e).

LDA calculated Fermi surface for the Ba system is quite simple with respect to
the shape of different sheets. That can be seen on the panel (a) of lower
row of the Fig.~4. Again the Fermi surface is essentially three-dimensional and
also has three sheets. In the centre there is large hole-like Fermi surface sheet
(see panel (b) in the lower line of Fig.~4) while in the corners
there are two electron-like sheets.

In conclusion we performed LDA calculations of electronic band structure for
recently reported  in Ref.~\cite{Anand} materials (Sr,Ca)Pd$_2$As$_2$ and BaPd$_2$As$_2$
related to the 122 Fe-pnictide systems. In general band structure of the Sr system
is very similar to that of isovalent material SrNi$_2$As$_2$ reported elsewhere \cite{Shein_Ni}.
However the band structure of another isovalent system BaNi$_2$As$_2$ is
surprisingly quite different to the SrPd$_2$As$_2$ compound though the crystal structure is very similar.
The band structure of BaPd$_2$As$_2$ system is very dissimilar to any other
122 systems because of its different crystal structure. 
It is interesting that considered in this work Sr and Ba systems has one similarity within
the band structure. Namely, in the DOS of both materials there is rather intensive
peak formed by Pd-4d($x^2-y^2$) states (about 0.7 eV below the Fermi level).
Thus doping of about 2 holes per unit cell may lead to a considerable increase (by a factor 2.5)
of the total DOS value at the Fermi level $N(E_F)$ giving rise to $T_c$ for the Sr compound
and perhaps appearing of superconductivity in the Ba compound (traces of superconductivity
were observed here experimentally in Ref.~\cite{Anand}).
Obtained in LDA calculations values of $N(E_F)$ for stoichoimetric systems
are in perfect agreement with experimental estimates \cite {Anand} thus signifying
minor role of correlation effects in these systems.

\begin{figure}
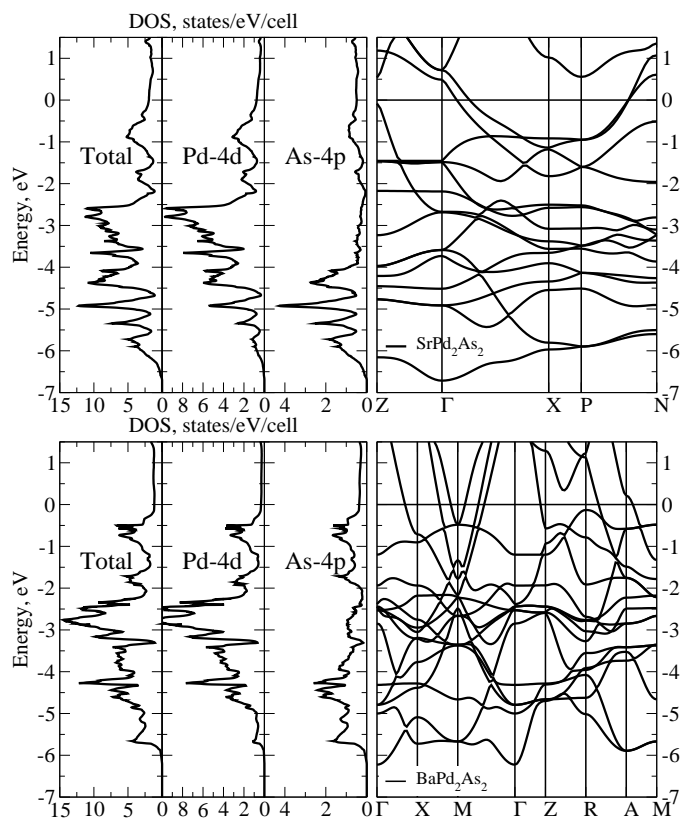

\includegraphics[clip=true,width=0.5\textwidth]{Sr_DOS_bands.eps}
\includegraphics[clip=true,width=0.5\textwidth]{Ba_DOS_bands.eps}
\caption{Fig. 2. LDA calculated band dispersions and densities of states
of (SrPd$_2$As$_2$ (top) and BaPd$_2$As$_2$ (bottom).
The Fermi level is zero.} 
\end{figure}

\begin{figure}
\includegraphics[clip=true,width=0.5\textwidth]{Sr_pdos.eps}
\includegraphics[clip=true,width=0.5\textwidth]{Ba_pdos.eps}
\caption{Fig. 3. Orbitally resolved densities of Pd-4d states from LDA calculations for 
(Sr,Ca)Pd$_2$As$_2$ (top) and BaPd$_2$As$_2$ (bottom).
The Fermi level is zero.} 
\end{figure}

\begin{figure*}
\begin{center}
\includegraphics[clip=true,width=1\textwidth]{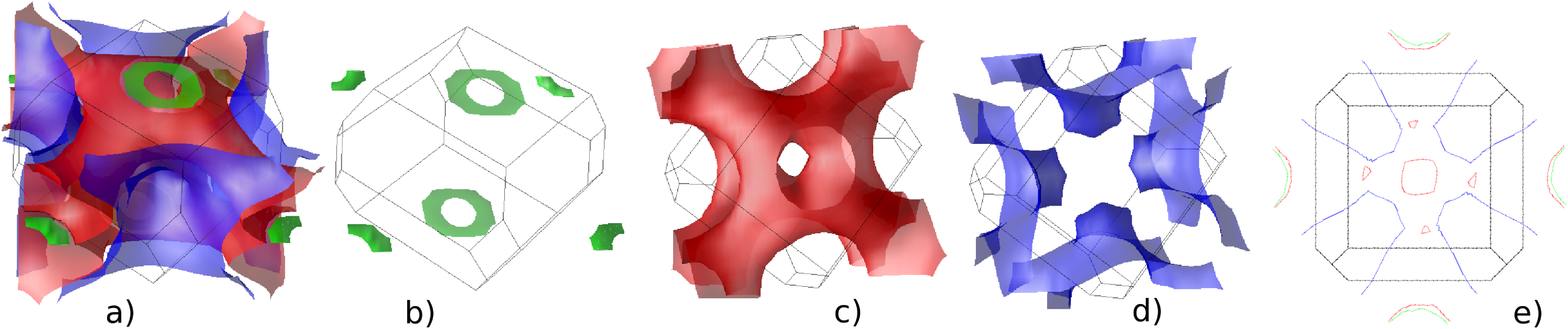}
\includegraphics[clip=true,width=0.6\textwidth]{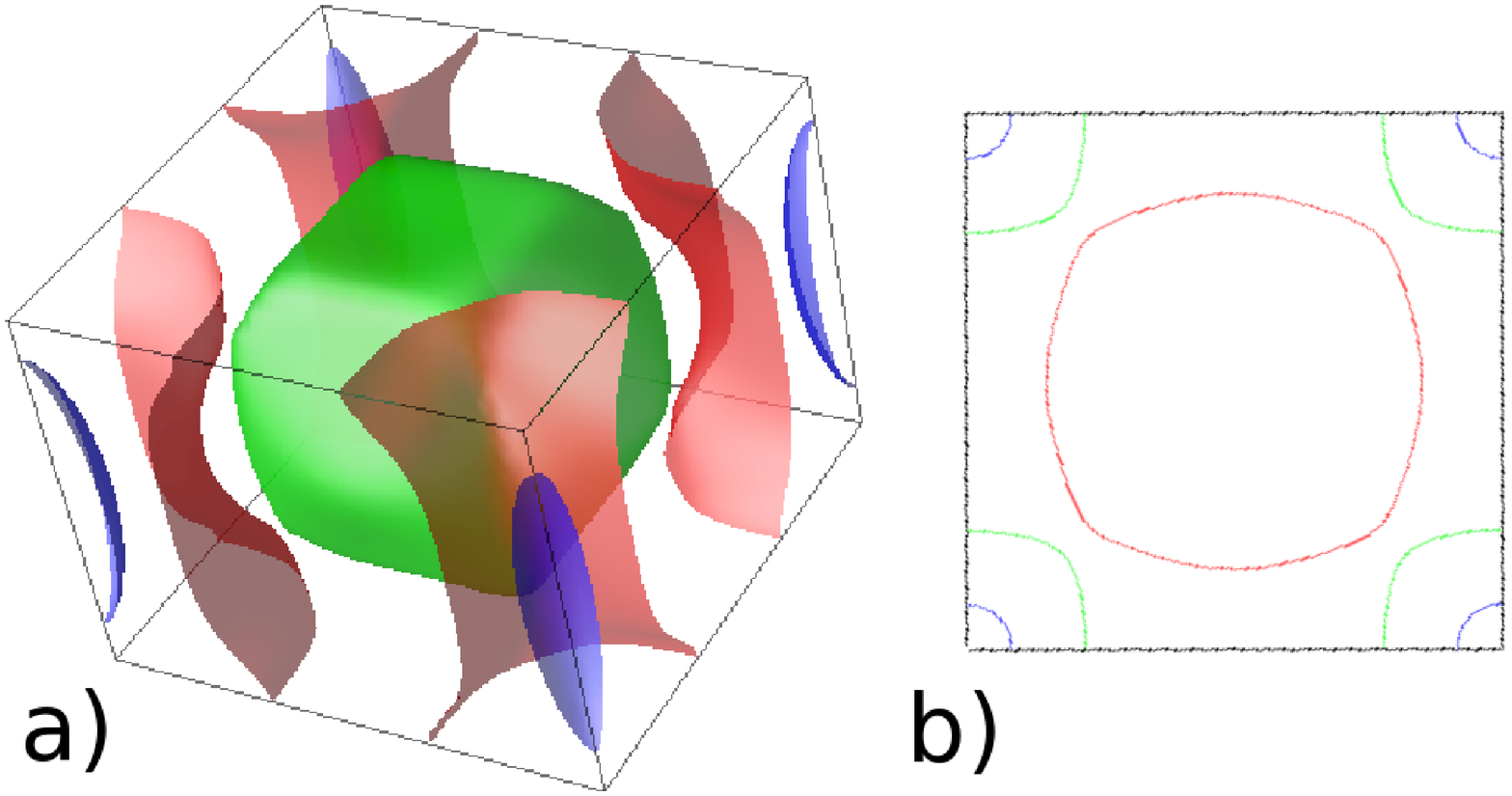}
\end{center}
\caption{Fig. 4. LDA calculated FS for (Sr,Ca)Pd$_2$As$_2$ (top) and BaPd$_2$As$_2$ (bottom).
a -- all FS sheets together for both systems;
b,c,d (top panel) -- separate view of each of three FS sheets for (Sr,Ca)Pd$_2$As$_2$;
e (top panel) and b (lower panel) -- crossection of FS at $k_z$=0  for  (Sr,Ca)Pd$_2$As$_2$ and BaPd$_2$As$_2$
correspondingly.} 
\end{figure*}

This work is partly supported by RFBR grant 11-02-00147 and was performed
within the framework of programs of fundamental research of the Russian
Academy of Sciences (RAS) ``Quantum mesoscopic and disordered structures''
(12-$\Pi$-2-1002). IAN acknowledges SB-UB RAS grant.

\end{document}